\documentclass[12pt]{book}

\usepackage{plenum}

\begin{document}

\title{CONFORMALITY, PARTICLE PHENOMENOLOGY AND THE COSMOLOGICAL CONSTANT}
\author{Paul H. Frampton}
\affiliation{Department of Physics and Astronomy, \\
University of North Carolina,\\
Chapel Hill, NC 27599-3255.}

\bigskip
\bigskip

\noindent {\bf Abstract}

\bigskip

\noindent Conformality is the idea that at TeV scales enrichment of the standard model
particle spectrum leads to conformal invariance at a fixed point
of the renormalization group. Some aspects of conformality 
in particle phenomenology and cosmology are discussed.

\bigskip

\section{Alternative to ``Grand'' Unification}

\bigskip

In GUT theories there is an unexplained hierarchy 
between the GUT scale and the weak scale which is 
about 14 orders of magnitude. There is the question 
of why these very different scales exist and how 
are the scales stabilized under quantum corrections? 

\bigskip

Supersymmetry solves the second of these problems but not the first.
Supersymmetry has some {\it successes}: 
(i) the cancellation of some UV divergences; 
(ii) the technical naturalness of the hierarchy; 
(iii) the unification of the gauge couplings; and
(iv) its natural appearance in string theory.

\bigskip

On the other side, supersymmetry definitely 
presents several {\it puzzles}:
(i) the ``mu'' problem - why is the Higgs at 
the weak scale not at the GUT scale?; 
(ii) breaking supersymmetry leads to too large a cosmological
constant; and (iii) is supersymmetry really fundamental 
for string theory since there are solutions of string theory without supersymmetry.

\bigskip

These general considerations led naturally to the 
suggestion [\cite{PHF1,PHF2,PHF3,PHF4,PHF5,PHF6}]
that supersymmetry and grand unification 
should be replaced by conformality at the TeV scale.
Here it will be shown that this idea is possible, 
including explicit examples containing the standard 
model states. Further it will be shown that conformality 
is a much more rigid constraint than supersymmetry.
Conformality predicts additional states at the TeV scale 
and a rich inter-family structure of Yukawa couplings.

\bigskip
\bigskip

\section{Conformality as a Hierarchy Solution}

\bigskip

First we note that quark and lepton masses, the QCD scale 
and weak scale are small compared to a (multi-) TeV scale.
At the higher scale they may be put to zero, suggesting 
the addition of further degrees of freedom
to yield a quantum field theory with conformal invariance.
This has the virtue of possessing 
naturalness in the sense of 't Hooft [\cite{Hooft}] since
zero masses and scales increases the symmetry.

\bigskip

The theory is assumed to be given by the action:

\begin{equation}
S = S_0 + \int d^4 x \alpha_i O_i
\end{equation}
where $S_0$ is the action for the conformal theory and the $O_i$
are operators with dimension below four which break conformal invariance softly.

\bigskip

\noindent The mass parameters $\alpha_i$ have 
mass dimension $4 - \Delta_i$ where $\Delta_i$ is the dimension of $O_i$
at the conformal point.

\bigskip

\noindent Let $M$ be the scale set by the parameters 
$\alpha_i$ and hence the scale at which conformal invariance is
broken. The for $E \gg M$ the couplings will not run
while they start running for $E < M$. To solve the hierarchy 
problem we assume $M$ is near to the TeV scale.

\bigskip
\bigskip

\section{d = 4 CFTs}

\bigskip 

In enumerating the CFTs in 4 spacetime dimensions, we must choose
the $N$ of $SU(N)$. To leading order in $1/N$, the RG $\beta$-functions
always vanish as they coincide with the ${\cal N}=4$ case [\cite{Vafa1,Vafa2}].
For finite $N$ the situation is still under active investigation. 
To {\it prove} the $\beta-$ functions vanish when ${\cal N} =0$
is rendered more difficult by the fact that without
supersymmetry the associated nonrenormalization theorems are absent.

We extract the candidates from compactification[\cite{maldacena}]
of the Type IIB superstring on $AdS_5 \times S^5/\Gamma$.

Let $\Gamma\subset SU(4)$ denote a discrete subgroup of $SU(4)$.
Consider irreducible representations of $\Gamma$.  Suppose there
are $k$ irreducible representations $R_i$, with dimensions $d_i$ with
$i=1,...,k$.  The gauge theory in question has gauge symmetry

\begin{equation}
SU(N d_1)\times SU(Nd_2)\times ...SU(Nd_k)
\end{equation}

The fermions in the theory are given as follows.  
Consider the 4 dimensional
representation of $\Gamma$ induced from its embedding in $SU(4)$.  It
may or may not be an irreducible representation of $\Gamma$. We consider
the tensor product of ${\bf 4}$ with the representations $R_i$:

\begin{equation}
{\bf 4}\otimes R_i=\oplus_j n_i^jR_j
\end{equation}

The chiral fermions are in bifundamental representations

\begin{equation}
(1,1,..,{\bf Nd_i},1,...,{\overline {{\bf Nd_j}}},1,..)
\end{equation}
with multiplicity $n_i^j$ defined above.  For $i=j$ the
above is understood in the sense that we obtain
$n_i^i$  adjoint fields plus $n_i^i$ singlet 
fields of $SU(Nd_i)$.

\bigskip

Note that we can equivalently view
$n_i^j$ as the number of trivial representations in the tensor product
\begin{equation}
({\bf 4}\otimes R_i\otimes R_j^*)_{trivial}=n_i^j
\end{equation}

The asymmetry between $i$ and $j$ is manifest in the above
formula. Thus in general we have
$n_i^j\not= n_j^i$ and so the theory in question
is in general a chiral theory.  However
if $\Gamma$ is a real
subgroup of $SU(4)$, i.e. if ${\bf 4}={\bf 4}^*$ as far as
$\Gamma$ representations are concerned,
then we have by taking the complex
conjugate:

\begin{equation}
n_i^j=({\bf 4}\otimes R_i\otimes R_j^*)_{trivial}=
({\bf 4}\otimes R_i\otimes R_j^*)^*_{trivial}=
({\bf 4}^*\otimes R_i^*\otimes R_j)_{trivial}=
({\bf 4} \otimes R_i^*\otimes R_j)_{trivial}=n_j^i.
\end{equation}

So the theory is chiral only if ${\bf 4}$ is a complex
representation of $\Gamma$, i.e. only if ${\bf 4}\not={\bf 4}^*$
as a representation of $\Gamma$.
If $\Gamma$ were a real subgroup of $SU(4)$ then
$n_i^j=n_j^i$.

If $\Gamma$ is a complex subgroup, the theory is chiral, but
it is free of gauge anomalies.  To see this note that
the number of chiral fermions in the
fundamental representation of each group $SU(Nd_i)$ plus $Nd_i$
times the number of chiral fermions in the 
adoint representation is given by

\begin{equation}
\sum_j n_i^j Nd_j=4 Nd_i
\label{sum}
\end{equation}
(where the number of adjoints is given by $n_i^i$).
Similarly the number of anti-fundamentals plus $Nd_i$ times
the number of adjoints is given by

\begin{equation}
\sum_j n_j^i Nd_j=\sum Nd_j(4\otimes R_j\otimes R_i^*)_{trivi
al}=
\sum Nd_j(4^*\otimes R_j^*\otimes R_i)_{trivial}=4 Nd_i
\end{equation}

Thus, comparing with Eq.(\ref{sum}) we see that
the difference of the number of chiral fermions
in the fundamental and the antifundamental representation
is zero (note that the adjoint representation is real and does
not contribute to anomaly). Thus each gauge group is anomaly free.
The requirement of anomaly cancellation is, of course, a familiar
one in string theory [\cite{FK1,FK2}] as well as in model
building beyond the standard model [\cite{FG1,FG2,PHF,FN}].

\bigskip
\bigskip

In addition to fermions, we have bosons, also in 
the bifundamental
represenations.  The number of bosons $M_i^j$
in the bifundamental representation
of $SU(Nd_i)\otimes SU(Nd_j)$ is 
given by the number of $R_j$ representations
in the tensor product of the representation ${\bf 6}$ of
$SU(4)$
restricted to $\Gamma$ with the 
$R_i$ representation.  Note that
since ${\bf 6}$ is a real representation we have
$$M_i^j=(6\otimes R_i\otimes R_j^*)_{trivial}=(6\otimes R
_i^*\otimes
R_j)_{trivial}=M_j^i$$
In other words for each $M_i^j$ 
we have a {\it complex} scalar
field in the corresponding bifundamental representation,
where complex
conjugation will take us from the 
fields labeled by $M_i^j$ to $M_j^i$.

The fields in the theory are naturally 
summarized by a graph, called
the quiver diagram [\cite{DM}], where 
for each gauge group $SU(Nd_i)$
there
corresponds a node in the graph, for each chiral fermion
in the
representation $(Nd_i, {\overline Nd_j})$, 
$n_i^j$ in total, corresponds a directed
arrow from the $i$-th node to the $j$-th node, 
and for each complex
scalar in the bifundamental of $SU(Nd_i)\times SU(Nd_j)$,
$M_i^j$ in
total, corresponds
an {\it undirected} line between the $i$-th node and the
$j$-th node

\bigskip
\bigskip

\noindent Interactions. Gauge fields interact 
according to gauge coupling which,
combined with corresponding theta 
angle for i th group, is writable as
\[
\tau_i = \Theta_i + \frac{i}{4\pi g_i^2} = \frac {\tau d_i}{|\Gamma|}
\]
where $\tau$ is complex parameter (independent i) and $|\Gamma|=$ order 
$\Gamma$.
\bigskip
\bigskip

\noindent Yukawa interactions. Triangles in quiver. Two directed fermion sides 
and an undirected scalar side.
\[
S_{Yukawa} = \frac{1}{4 \pi g^2} \sum d^{abc} Tr \Psi^a _{ij} \Phi^b_{jk} 
\Psi^c_{ki}
\]
in which $d^{abc}$ is ascertainable as Clebsch-Gordan coefficient from
product of trivial representaions occurring 
respectively in $(4 \bigotimes R_i \bigotimes  R_j^*)$,
$(6 \bigotimes R_j \bigotimes R_k^*)$ and $(4 \bigotimes R_k \bigotimes R_i^*)$.

\bigskip
\bigskip

\noindent Quartic scalar interactions. Quadrilaterals in quiver. Four undirected
sides. The coupling computable analagously to above.

\bigskip
\bigskip

\noindent Conformality. To leading order 
in 1/N all such theories are confromal[\cite{Vafa1,Vafa2}].

\bigskip

\noindent Are they conformal for higher orders?

\bigskip

\noindent YES, for ${\cal N} = 2$. All such ${\cal N} = 2$
theories are obtainable.

\bigskip

\noindent YES, for ${\cal N} = 1$: non-renormalization
theorems ensure flat direction(s).

\bigskip

\noindent UNKNOWN for ${\cal N} = 0$.

\bigskip
\bigskip

\noindent Conformality for ${\cal N} = 0$.
We can offer a plausibility argument 
for a conformal IR fixed point. If only
one independent coupling occurs then the 
S-duality of the progenitor Type IIB superstring 
implies $g\rightarrow1/g$ symmetry. 
If the next to leading order in
1/N is asymptotically free then IR flow 
increases g. Therefore for large g IR flow 
decreases g. Hence $\beta_g = 0$ 
for some intermediate g.

\bigskip
\bigskip

\section{Applications of Conformality to Particle Phenomenology.}

\bigskip

\noindent It is assumed that the Lagrangian is {\it nearly} conformal.
That is, it is the soft-breaking of a conformal theory.

\bigskip
\bigskip

\noindent The soft breaking terms would involve quadratic and cubic
scalar terms, and fermion mass terms. In the quiver diagram, these correspond
respectively to 2-gons
and triangles with undirected edges, and 2-gons
with compatibly directed edges.

\bigskip
\bigskip

\[
S = S_0 + \int \alpha_{ab} Tr \Psi^a_{1j*}\Psi^b_{ji*} + \alpha_{cd}^2 Tr 
\Phi^c_{1j*}
\Phi^d_{ji*}
\]
\[ + \alpha_{efg} Tr \Phi^c_{ij*}\Phi^f_{jk*}\Phi^g_{ki*} + c.c.
\]

\bigskip
\bigskip

\noindent Depending on the sign of the scalar mass term the conformal 
breaking could induce gauge symmetry breaking.

\bigskip
\bigskip

Consider a gauge subgroup 
$SU(Nd_i) \times SU(Nd_j)$ and suppose that $<\Phi_{ij*}> \neq 0$. Assume for 
simplicity that
$d_i=d_j=d$. Then the VEV can be represented by a square matrix with diagonal 
entries. The symmetry breaking depends on the eigenvalues. If there are two 
equal eigenvalues and the rest zero we get:
\[
SU(Nd) \times SU(Nd) \rightarrow 
\]
\[SU(2)_{diagonal} \times U(1) \times SU(Nd-2) \times
SU(Nd-2)
\]

\bigskip
\bigskip

\noindent
With more such VEVs and various alignments thereof a rich pattern of gauge
symmetry breakings can emerge.

\bigskip
\bigskip

\noindent GENERAL PREDICTIONS.

\bigskip
\bigskip

\noindent Consider embedding the standard 
model gauge group according to:
\[
SU(3) \times SU(2) \times U(1) \subset \bigotimes_i SU(Nd_i)
\]
Each gauge group of the SM can lie entirely in a $SU(Nd_i)$ 
or in a diagonal subgroup of a number thereof.

\bigskip

\noindent Only bifundamentals (including adjoints) are possible.
This implies no $(8,2)$, etc. A conformality restriction which is new and
satisfied in Nature!

\bigskip

\noindent No $U(1)$ factor can be conformal and so hypercharge is quantized
through its incorporation in a non-abelian gauge group.
This is the ``conformality'' equivalent to the GUT charge quantization
condition in {\it e.g.} $SU(5)$!

\bigskip

\noindent Beyond these general consistencies, there are predictions of
new particles
necessary to render the theory conformal. 

\bigskip
\bigskip

\noindent The minimal extra particle content 
comes from putting each SM gauge group in one
$SU(Nd_i)$.
Diagonal subgroup embedding {\it increases} 
number of additional states.

\bigskip

\noindent Number of fundamentals plus $Nd_i$ 
times the adjoints is $4Nd_i$.
Number $N_3$ of color triplets 
and $N_8$ of color octets satisfies:
\[
N_3 + 3N_8 \geq 4 \times 3 = 12
\]
Since the SM has $N_3=6$ we predict:
\[
\Delta N_3 + 3N_8 \geq 6
\]
The additional states are at TeV if conformality
solves hierarchy.
Similarly for color scalars:
\[
M_3 + 3M_8 \geq 6 \times 3 = 18
\]
The same exercise for $SU(2)$ gives 
$\Delta N_2 + 4N_3 \geq 4$ and 
$\Delta M_2 + 2M_3 \geq 11$ respectively.

\bigskip
\bigskip

\noindent FURTHER PREDICTIONS

\bigskip
\bigskip

Yukawa and Quartic interactions are untouched 
by soft-breaking terms. These are therefore completely 
determined by the IR fixed point parameters. 
So a rich structure for flavor is dictated by conformal invariance.
This is to be compared with the MSSM (or SM) where the 
Yukawa couplings are free parameters.     

\bigskip
\bigskip

\noindent GAUGE COUPLING UNIFICATION

\bigskip
\bigskip

\noindent Above the TeV scale couplings will not run. 
The couplings are nevertheless related, and not necessarily equal at the conformal scale.

\bigskip

\noindent For example, with equal $SU(Nd)$ couplings embed
$SU(3)$, $SU(2)$, and $U(1)$ diagonally into 
1, 3, 6 such groups respectively to obtain proximity to 
the correct ratios of the low-energy SM gauge couplings.

\bigskip
\bigskip

\noindent Some illustrative examples of model building using conformality:

\bigskip
\bigskip

\noindent We need to specify an embedding $\Gamma \subset SU(4)$.

\bigskip

\noindent Consider $Z_2$. It embeds as $(-1, -1, -1, -1)$ which
is real and so leads to a non-chiral model.

\bigskip

$Z_3$. One choice is {\bf 4} = $(\alpha, \alpha, \alpha, 1)$
which maintains {\cal N}=1 supersymmetry. 
Otherwise we may choose {\bf 4}=$(\alpha, \alpha, \alpha^2, \alpha^2)$
but this is real.

\bigskip

$Z_4$. The only {\cal N} = 0 complex embedding is
{\bf 4}=$(i, i, i, i)$. The quiver is as shown on the
next transparency with the $SU(N)^4$ gauge groups at the corners,
the fermions on the edges and the scalars on the diagonals.
The scalar content is too tight to break to the SM.

\bigskip
\bigskip

Naming the nodes 0, 1, 2, 3, 4 we identify 0 with color
and the diagonal subgroups (1,3) and (2,4) with
weak and hypercolor respectively. There are then
three families in
\[
(3, \bar{3}, 1) + (1, 3, \bar{3}) + (\bar{3}, 1, 3)
\]
and one anti-family.

\bigskip
\bigskip

\noindent We suppose that the soft conformal
breaking excludes a mass term marrying the third family to
its mirror.

\bigskip
\bigskip

\noindent There are sufficient scalars to break to
the SM with three families.

\bigskip
\bigskip

\noindent This is an existence proof.

\bigskip
\bigskip

\noindent Simplest three family model has ${\cal N}=1$
supersymmetry.

\bigskip
\bigskip

\[
Z_3.~~~ 4= (\alpha, \alpha, \alpha, 1)
\]

\bigskip
\bigskip

\noindent Fermions and scalars are:

\[
\sum_{i=1}^3 (3_i, \bar{3}_{i+1}) + \sum_{i=1}^3(8+1)_i
\]

\bigskip
\bigskip

\[
\beta_g^{(1)} = -\frac{g^3}{16 \pi^2} \left[ \frac{11}{3}C_2(G)
-\frac{2}{3}T(R_W) - \frac{1}{6}T(R_R) \right]
\]

\bigskip
\bigskip

\noindent Find:

\[\beta_g^{(1)} \sim 9 - 9 = 0
\]

\bigskip
\bigskip

\noindent for all three $SU(3)$ factors in
supersymmetric trinification.

\bigskip
\bigskip

\noindent NON-ABELIAN ORBIFOLDS

\bigskip

We consider all non-abelian discrete groups up to order $g < 32$.
There are exactly 45 such groups.
Because the gauge group arrived at is $\otimes_i SU(Nd_i)$ we
can arrive at $SU(4)\times SU(2) \times SU(2)$ by choosing 
$N = 2$.

\bigskip

To obtain chiral fermions one must have {\bf 4} $\neq$ {\bf 4$^*$}
This is not quite sufficient because for $N = 2$, if {\bf 4} 
is complex but pseudoreal, the fermions are still non-chiral [\cite{PHF6}].

\bigskip

This last requirement eliminates many of the 45 candidate groups.
For example $Q_{2n} \subset SU(2)$ has irreps of
appropriate dimensions but cannot sustain chiral fermions.
because these irreps are , like $SU(2)$, pseudoreal.

\bigskip

This leaves 19 possible non-abelian $\Gamma$ with $g \leq 31$,
the lowest order being $g =16$. This gives only two families.

\bigskip

The smallest group which allows three chiral familes has order $g = 24$
so we now describe this model.

\bigskip
\bigskip

Using only $D_N, ~~Q_{2N}, ~~S_N ~~{\rm and}~~ T$ 

(T = tetrahedral~~$S_4/Z_2$)
one already finds 32 of the 45 non-abelian discrete groups with $g \leq 31$:

\bigskip
\bigskip

\begin{tabular}{| | c | l | |}   \hline\hline
g  &   \\   \hline\hline 
6 &  $D_3 \equiv S_3$ \\ \hline
8 &  $D_4,~~Q = Q_4$ \\  \hline
10 & $ D_5$ \\  \hline
12 & $ D_6,~~Q_6,~~T$ \\  \hline
14 & $ D_7$ \\  \hline
16 & $ D_8,~~Q_8,~~Z_2 \times D_4,~~Z_2 \times Q $\\ \hline
18 & $ D_9,~~Z_3 \times D_3$ \\  \hline
20 & $ D_{10},~~Q_{10} $\\ \hline
22 & $ D_{11} $ \\  \hline
24 & $ D_{12},~~Q_{12},~~Z_2 \times D_6,
~~Z_2 \times Q_6,~~Z_2 \times T $ \\
&  $ Z_3 \times D_4,~~Z_3 \times Q,
~~Z_4 \times D_3,~~S_4$  \\ \hline
26  &  $D_{13}$ \\ \hline
28  &  $D_{14},~~Q_{14} $ \\ \hline
30  &  $D_{15},~~D_5 \times Z_3,~~D_3 \times Z_5$  \\  \hline\hline
\end{tabular}

\bigskip

\noindent The remaining 13 of the 45 
non-abelian discrete groups with $g \leq 31$
are twisted products:

\bigskip

\begin{tabular}{| | c | l | |}   \hline\hline
g  &   \\   \hline\hline 
16 &  $Z_2 \tilde{\times} Z_8 ({\rm two, excluding} ~D_8),
~~Z_4 \tilde{\times} Z_4 $\\
&  $Z_2 \tilde{\times} (Z_2 \times Z_4) ({\rm two})$ \\ \hline
18 & $  Z_2 \tilde{\times} (Z_3 \times Z_3) $\\ \hline
20 & $ Z_4 \tilde{\times} Z_5 $ \\  \hline
21 & $ Z_3 \tilde{\times} Z_7 $ \\  \hline
24  &$  Z_3 \tilde{\times}  Q,~~Z_3 \tilde{\times} Z_8,
~~Z_3 \tilde{\times} D_4$  \\ \hline
27 &  $ Z_9 \tilde{\times} Z_3,
~~Z_3 \tilde{\times} (Z_3 \times Z_3)$ \\  \hline\hline
\end{tabular}

\bigskip
\bigskip

Successful $g = 24$ model is based on the group
$\Gamma = Z_3 \times  Q$. 

\noindent The fifteen irreps of $\Gamma$ are 

\noindent $1$,~$1'$,~$1''$,~$1'''$,~$2$,

\noindent ~$1\alpha$, ~$1'\alpha$,~$1''\alpha$,~$1'''\alpha$,~$2\alpha$,

\noindent 
$1\alpha^{-1}$,~$1'\alpha^{-1}$,~$1''\alpha^{-1}$,~$1'''\alpha^{-1}$,~$2\alpha^{
-1}$.

\noindent The same model occurs for $\Gamma = Z_3 \times D_4$. The 
multiplication table
is shown below.

\bigskip
\bigskip

\begin{tabular}{| | c | | c | c | c | c | c | |}\hline\hline
 & 1 & $1'$ & $1''$ & $1'''$ & 2  \\    \hline\hline 
1 & 1 & $1'$ & $1''$ & $1'''$ & 2 \\  \hline
$1'$ & $1'$ & $1$ & $1'''$ & $1''$ & $2$ \\ \hline
$1''$ & $1''$ & $1'''$ & $1$ &$1'$ & $2$ \\ \hline
$1'''$ & $1'''$ & $1''$ & $1'$ & $1$  & $2$  \\ \hline
$2$   & $2$ & $2$ & $2$ & $2$ &  $1+1'$ \\ 
    &    &    &    &    &   $1''+1'''$ \\ \hline\hline
$1\alpha$ & $1\alpha$ & $1'\alpha$ & $1''\alpha$ & $1'''\alpha$ & $2\alpha$ \\  
\hline
$1'\alpha$ & $1'\alpha$ &$1\alpha$ & $1'''\alpha$ & $1''\alpha$ & $2\alpha$ \\ 
\hline
$1''\alpha$ & $1''\alpha$ & $1'''\alpha$ & $1\alpha$ & $1'\alpha$ & $2\alpha$ \\
\hline
$1'''\alpha$ & $1'''\alpha$ & $1''\alpha$ & $1'\alpha$ & $1\alpha$  & $2\alpha$ 
\\ \hline
$2\alpha$   & $2\alpha$ & $2\alpha$ & $2\alpha$ & $2\alpha$ &  
$1\alpha+1'\alpha$ \\ 
    &    &    &    &    &   $1''\alpha+1'''\alpha$ \\ \hline\hline
\end{tabular}

\bigskip
\bigskip

etc.

\bigskip
\bigskip

The general embedding of the required type can be written:

\bigskip

{\bf 4} = $(1\alpha^{a_1},~1'\alpha^{a_2},~2\alpha^{a_3})$  

\bigskip

The requirement that the {\bf 6} is real dictates that

\bigskip

$a_1 + a_2 = - 2a_3$

\bigskip

It is therefore sufficient to consider for ${\cal N}$ = 0 no surviving
supersymmetry only the choice:

\bigskip

{\bf 4} = $(~1\alpha, ~1', ~2\alpha)$

\bigskip

It remains to derive the chiral fermions and the complex scalars
using the procedures already discussed (quiver diagrams).

\bigskip
\bigskip

\noindent \underline {$D_4 \times Z_3$ MODEL.}

\bigskip
\bigskip

\noindent VEVs for these scalars allow to break to the 

\noindent following diagonal subgroups as the only 

\noindent surviving gauge symmetries:

\bigskip

$SU(2)_{1,2,3} \longrightarrow SU(2)$

\bigskip

$SU(2)_{5,6,7} \longrightarrow SU(2)$

\bigskip
\bigskip

$SU(4)_{1,2} \longrightarrow SU(4)$

\bigskip
\bigskip

\noindent This spontaneous symmetry breaking leaves the Pati-Salam type
model:

\bigskip
\bigskip

$SU(4) \times SU(2) \times SU(2)$

\bigskip
\bigskip

\noindent with three chiral fermion generations

\bigskip
\bigskip

$3~[(~4,~2,~2) + (~\bar{4},~2,~2)]$

\bigskip
\bigskip

\section{Towards the Cosmological Constant.}

\bigskip
\bigskip

\noindent INCLUSION OF GRAVITY.

\bigskip
\bigskip

\noindent The CFT arrived at is in a flat spacetime background which does not 
contain gravity.

\bigskip
\bigskip

\noindent One way to introduce the four-dimensional 
graviton introduces an extra
spacetime dimension and truncates the range of the fifth dimension. 
The four-dimensional graviton then appears  by compactification of the
higher-dimensional graviton, as is certainly the path 
suggested by the superstring.   

\bigskip
\bigskip

\noindent Although conformality solves 
the hierarchy between the weak scale and the
GUT scale, the hierarchy existing in non-string theory without gravity,
it is clear that classical gravity violates conformal invariance
because of its dimensional Newton coupling constant.
The inclusion of gravity in the conformality scheme most likely
involves a change in the spacetime at the Planck scale; one
possibility being explored is noncommutative
spacetime coordinates [\cite{Seiberg}]. Another even more radical
idea is the one already mentioned to invoke [\cite{antoniadis}]
at TeV scales an extra spacetime coordinate.

\bigskip
\bigskip

\noindent SUMMARY.

\bigskip
\bigskip

\noindent Conformality is seen to be a rigid organizing principle. 
Many embeddings remain to be studied. Soft breaking of conformal 
symmetry deserves further study, as does the even more appealing 
case of spontaneous breaking of conformal symmetry.

\bigskip

\noindent The latter could entail flat directions 
even in the absence of supersymmetry
and if this is really possible one would need to invoke a symmetry
different from supersymmetry to
generate the flat direction.

\bigskip

\noindent This would lead naturally 
to an explanation of the vanishing
cosmological constant different from any where a
fifth spacetime dimension is invoked [\cite{KSS,ADKS}].

\bigskip
\bigskip

\noindent New particles await discovery at the 
TeV scale if the conformality
idea is valid.

\bigskip
\bigskip
\bigskip
\bigskip
\bigskip
\bigskip

\section{Acknowledgement.}

\bigskip

\noindent This work was supported in part by the 
US Department of Energy
under Grant No. DE-FG02-97ER41036.

\end{document}